  \documentclass[pra,10pt]{revtex4}

 \usepackage{amsmath} 
 \usepackage{graphicx}
 \usepackage{hyperref}
\hypersetup{colorlinks=true, linkcolor=blue, citecolor=blue, urlcolor=blue}

\begin{document}

 \def\BE{\begin{equation}}
 \def\EE{\end{equation}}
 \def\BA{\begin{array}}
 \def\EA{\end{array}}
 \def\BEA{\begin{eqnarray}}
 \def\EEA{\end{eqnarray}}
 \def\nn{\nonumber}
 \def\ra{\rangle}
 \def\la{\langle}

  \title{Schr{\"o}dinger cat states prepared by logical gate with non-Gaussian resource state: 
effect of finite squeezing and efficiency versus monotones}
 \author{A.~V.~Baeva}
 \author{I.~V.~Sokolov}
 \email{i.sokolov@spbu.ru, sokolov.i.v@gmail.com}
 \affiliation{Saint Petersburg State University, Universitetskaya nab.  7/9, 199034  Saint Petersburg, Russia}

\begin{abstract}
Quantum measurement-induced gate based on entanglement with ideal cubic phase state used as a non-Gaussian resource 
is able to produce Shr{\"o}dinger cat state in the form of two high fidelity ``copies'' of the 
target state on phase plane [N.I. Masalaeva, I.V. Sokolov, Phys. Lett. A {\bf 424}, 127846 (2022)]. 
In this work we examine the effect of finite initial squeezing of the resource state on the gate performance. 
We present exact solution for the gate output state and demonstrate that there exists a degree of squeezing, 
available in experiment, such that the output cat state quality almost does not impove with the further increase of squeezing. 
On the other hand, the probability of the expected ancilla measurement outcome decreases with squeezing. 
Since an overall efficiency of the conditional scheme should account for the probability of success, we
argue that such measures of non-Gaussianity of the resource state, as Wigner logarithmic negativiy and 
non-Gaussianity, may not be directly applicable to assess the efficiency of non-Gaussian gates based 
on quantum entanglement and subsequent projective measurement.
\end{abstract}


 \maketitle

 \section{Introduction}

The continuous-variable (CV) quantum information schemes based on Gaussian resource states were extensively explored 
both theoretically and experimentally~\cite{Lloyd99,Braunstein05,Furusawa98,Li02}, including their essentially multimode 
implementation~\cite{Gu09,Weedbrook12,Ukai15,Yokoyama13,Roslund14}. While the continuous-variable Gaussian cluster schemes 
are able to perform Gaussian transformations of the input states, in order to achieve universal quantum computing 
there is a need to introduce the non-Gaussian logical gates ~\cite{Lloyd99,Braunstein05}.

A minimal nonlinearity sufficient to prepare non-Gaussian resource states is cubic. 
The cubic phase state based on cubic nonlinearity was first considered in~\cite{Gottesman01,Bartlett_Sanders02}, 
and some approaches to the implementation of such states, as well as of the cubic (or higher) phase gates, were explored 
both theoretically and experimentally~\cite{Ghose07,Marek11,Yukawa13,Marshall15,Miyata16,Marek18}.

Recently, we demonstrated~\cite{Sokolov20,Masalaeva22} that continuous-variable measurement-induced two-node logical gate 
is able to prepare Schr{\"o}dinger cat-like quantum superpositions, if Gaussian (e. g. squeezed) resource state of an ancillary oscillator 
is substituted with non-Gaussian one, and standard homodyne measurement is applied. 
This was shown for the cubic phase state used as a resource. 

A variety of methods to prepare CV Schr{\"o}dinger cat states were discussed and implemented experimentally so far.
This can be directly achieved by unitary evolution assisted by a strong enough non-linear interaction~\cite{Yurke86}. 
The schemes based on a hybrid measurement-induced evolution also can create cat-like states.
The optical Schr{\"o}dinger cat states were generated using photon subtraction in a low-photon regime
\cite{Ourjoumtsev06} and from wide-band CV squeezed light \cite{Takase22},
the homodyne detection with photon number state as a resource~\cite{Ourjoumtsev07}, and
the iterative schemes which allow an incremental enlargement of cat states \cite{Etesse14,Etesse15,Sychev17}.
The Schr{\"o}dinger cat states can be prepared by using a photonic even-parity detector and CV entanglement
\cite{Thekaddath20}, by photon number measurement on two-mode  \cite{Takase21} or multimode \cite{Su19} Gaussian state.

A key feature of the gate~\cite{Sokolov20,Masalaeva22} is that the Schr{\"o}dinger cat state emerges when the ancillary oscillator 
measurement is compatible not with one, but with two different values of its physical variables. 
As far as these values are imprinted by the entanglement into the target oscillator observables, the two-component
cat-like quantum superposition is created at the output of the gate. This feature does not appear if Gaussian (squeezed) 
resource state is used, and needs the measurement which is consistent with this criterion.

Since the experimentally attainable  degree of squeezing is limited, in this work we focus on the effect of finite initial squeezing 
of the resource state on the gate performance. The exact solution for the gate output state in terms of the Airy function 
is presented for vacuum input state. We  demonstrate that there exists a degree of squeezing, 
available in experiment, such that the output superposition quality almost does not impove with the further increase of squeezing. 
This quality is estimated by means of the fidelity between the output state and the superposition of two coherent states
well spaced on the phase plane, which is of interest for some error correction protocols.

On the other hand, we find that the probability of the expected ancilla measurement outcome decreases with squeezing.
It is natural to admit that an overall efficiency of the conditional scheme should account for the probability of success. We
argue that such measures of non-Gaussianity of the resource state, as Wigner logarithmic negativiy and 
non-Gaussianity, may not be directly applicable to assess the efficiency of non-Gaussian gates, which are based 
on quantum entanglement and subsequent projective measurement.


 \section{Cat states from CV gate with cubic resource state based on finite initial squeezing}

The non-Gaussian gate based on perfect cubic phase state was introduced in~\cite{Sokolov20, Masalaeva22}. 
The measurement-induced two-node gate uses the cubic phase state of the ancillary oscillator as an elementary non-Gaussian 
resource, the $C_Z$ operation which entangles an input signal with the ancilla, and the projecting homodyne measurement.
Under optimal conditions, the gate output state is close to  ``perfect'' Schr{\"o}dinger cat state, that is,
to the superposition of two symmetrically displaced undistorted copies of the input state.

A key feature of the gate is that the ancilla measurement outcome provides multivalued information about the 
output state canonical variables, which results in the preparation of a cat-like state. 
This feature is easily interpreted \cite{Sokolov20, Masalaeva22} in terms of a clear pictorial representation which is extendable
to some other measurement-induced schemes based on the non-Gaussian resources.

Let us outline briefly the gate operation in a more realistic configuration, when the  
ancillary resource oscillator is subject to a finite initial squeezing before the cubic nonlinear perturbation is applied to 
the ancilla. In general, the target oscillator may be initially prepared in an arbitrary state
 $$
|\psi_1\ra = \int dx_1\psi(x_1)|x_1\ra,
 $$
which is assumed to occupy a limited range $\{\Delta x_1,\,\Delta y_1\}$ of the coordinate and momentum.

As the  initial state of the ancillary oscillator, prepared before the cubic evolution is applied, we consider the squeezed 
state whose uncertainty region on the phase plane is squeezed along the momentum axis by the factor $s \leq 1$,
 \BE
 \label{ancilla_sq}
\psi^{(sq)}(x_2) =  \frac{\sqrt{s}}{\pi^{1/4}}e^{-(sx_2)^{2}/2}.
 \EE
In order to prepare the non-Gaussian cubic resource state of ancilla, one applies the unitary evolution operator
$\exp(i\gamma q_2^3)$ to the state (\ref{ancilla_sq}).
In the following, we use the notation $\{q,\,p\}$  for the coordinate and momentum operators.

Next, the $C_Z$ entangling unitary evolution operator $\exp(iq_1q_2)$ is applied, which prepares the state
 %
 \BE
 |\psi_{12}\ra = \frac{\sqrt{s}}{\pi^{1/4}}
 \int dx_1dx_2\psi(x_1)  e^{ix_2(x_1 + \gamma x_2^2)}e^{-(sx_2)^{2}/2}|x_1\ra|x_2\ra,
 \label{entangled}
 \EE
 %
and the ancillary oscillator momentum is measured with the outcome $y_m$.

Projecting the state~\eqref{entangled} on the homodyne detector eigenstate $|y_m\ra_{p_2}$, one arrives at the 
target oscillator output state wave function (unnormalized),
 \BE
 \label{out_reduced}
\tilde\psi^{(out)}(x) = \psi_{1}(x)\tilde \varphi(x-y_m),
 \EE
where the target oscillator input state wave function is multiplied by the factor which is expressed \cite{Vallee} 
in terms of the Airy function,
 \begin{multline}
 \label{added}
\tilde \varphi(x-y_m) =
\frac{\sqrt{s}}{\pi^{3/4}\sqrt{2}}
\int dx' \; e^{ix'(x-y_{m}+\gamma x'^{2})}e^{-(sx')^{2}/2} =\\
\frac{\sqrt{2s}\,\pi^{1/4}}{(3\gamma)^{1/3}}
\exp\left[\frac{s^2}{6\gamma}\left(x - y_m + \frac{s^4}{18\gamma}\right) \right]
{\rm Ai}\left[\frac{1}{(3\gamma)^{1/3}}\left(x -y_m + \frac{s^4}{12\gamma}\right)\right].
 \end{multline}

The probability density of the ancilla momentum measurement outcome $y_m$ is
 \BE
P(y_m) = \la\tilde\psi^{(out)}|\tilde\psi^{(out)}\ra = \int dx  \big|\tilde\psi^{(out)}(x)\big|^2.
 \EE
The state conditionally prepared by the gate finally is given by 
 \BE
 \label{out}
\psi^{(out)}(x) = \frac{1}{\sqrt{P(y_m)}} \tilde\psi^{(out)}(x).
 \EE

Previously \cite{Masalaeva22}, we have explored in detail the performance of the gate based on the ideal initial
resource state (that is, in the limit where instead of the squeezed state (\ref{ancilla_sq}) one uses the 
momentum eigenstate $|0\ra_{p_2}$)
for a representative set of Fock states of the target oscillator. Here we focus on the effect of finite initial squeezing.
To be specific, we assume the target oscillator to be in the vacuum state which is given by (\ref{ancilla_sq}),
where $x_2\to x_1$, $s=1$.

As seen from the exact result (\ref{added}), in the limit of perfect squeezing, $s\to 0$, the added factor vanishes, 
and $P(y_m)\to 0$.
In this limit, the probability density to observe any given value $y_m$ of the ancilla momentum also vanishes. 
Therefore, in our previous work we were able to explore in detail quantum statistics of the emerging Schr{\"o}dinger 
cat-like states in dependence on the cubic interaction parameter $\gamma$ and on the measurement outcome 
$y_m$, but could not discuss the probability density and propose an optimal choice of the initial  ancilla
squeezing.

Let us first consider how a finite initial squeezing affects the quality of the emerging superpositions.

In \cite{Masalaeva22} we compared the exact solution obtained under the assumption of perfect squeezing,
and the state closest to it, which corresponds to the Schr{\"o}dinger cat state in the form of a superposition 
of two undistorted copies of the vacuum state of the target oscillator (i. e., the Glauber states), symmetrically 
spaced on the phase plane along the momentum quadrature,
 \BE
 \label{cat_Glauber_new}
|\psi_{cat}^{(out)}\ra = \frac{\left(e^{i\theta} |\alpha\ra + e^{-i\theta}|-\alpha\ra\right)}{\sqrt{2\big(1 + 
\cos(2\theta) e^{-2|\alpha|^2}\big)}},
 \EE
where
 \BE
 \label{theta_new}
\alpha = ip^{(+)},  \quad  p^{(+)} = \sqrt{y_m/3\gamma}, \quad \theta =
\frac{\pi}{4} - \frac{2}{3\sqrt{3\gamma}}y_m^{3/2}.
 \EE

We have shown that in the limit of perfect squeezing and within a relevant range of the parameters $\{\gamma, y_m\}$,
the proposed gate  prepares with high fidelity  the superpositions (\ref{cat_Glauber_new}) of Glauber states. 
Since the experimentally available squeezing  is limited, it remained unclear to which degree a finite squeezing 
can affect the possibility to obtain the Schr{\"o}dinger cat states close to (\ref{cat_Glauber_new}).

 \begin{figure}[h]
 \begin{center}
 \includegraphics[width=0.45\textwidth]{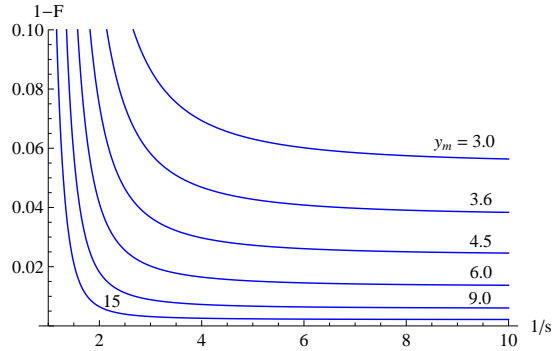}
 \caption{Infidelity $1-F_{cat}$ for the ancilla momentum measurement outcome $y_m=3,\,3.6,\,4.5,\,6,\,9,\,15$ 
 respectively  in dependence on the  initial ancilla state squeezing, where $1/s \geq 1$ is the stretching factor 
 of the ancilla coordinate  quadrature.  
 In these plots, the cubic deformation coefficient is chosen as $\gamma = y_m/30$, which provides
 the cat states with the same spacing $2p^{(+)} = 3.16$  (see (\ref{theta_new})) 
 between the copies along the momentum axis. }
 \label{fig_Infidelity}
 \end{center}
 \end{figure}

In Fig.~\ref{fig_Infidelity} we represent the effect of initial squeezing on the infidelity $1-F_{cat}$ between the gate
output state~\eqref{out} and the cat state~\eqref{cat_Glauber_new}, where
 \BE
 \label{cat_fidelity}
F_{cat} = \left| \int dx \psi^{(out)*}(x)\psi^{(out)}_{cat}(x)\right|^2.
 \EE

\begin{figure*}[h]
 \begin{center}
 \includegraphics[width=0.9\textwidth]{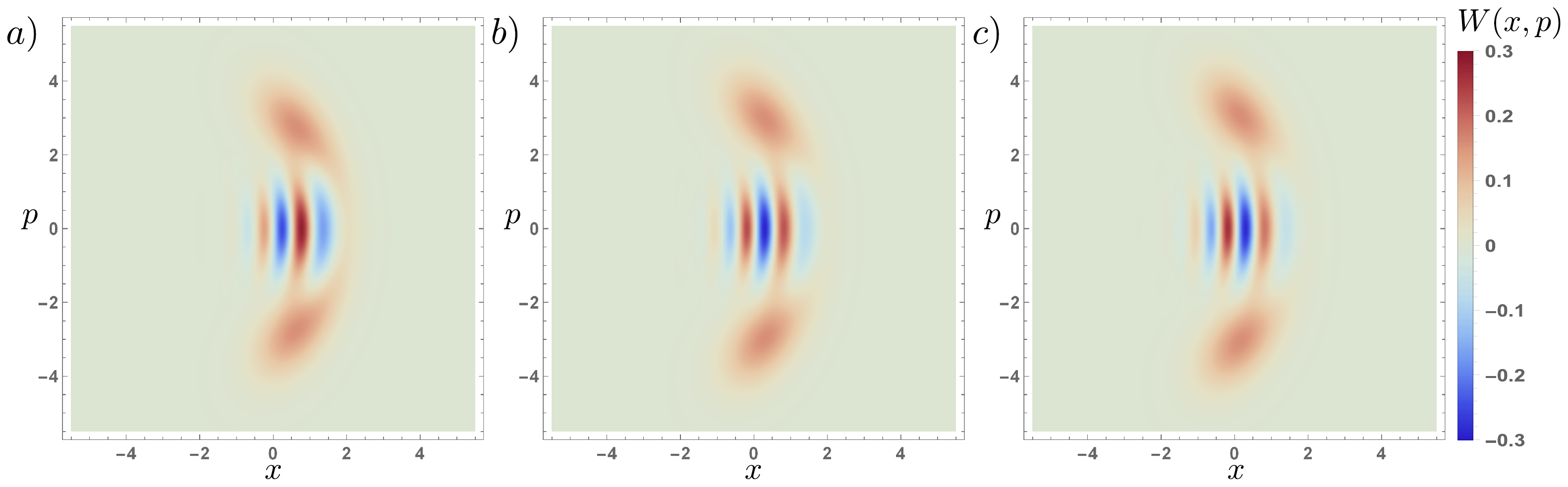}
 \caption{
Output state Wigner function for the cubic deformation coefficient $\gamma=0.1$ and the ancilla
momentum measurement outcome $y_m=3$ (low fidelity configuration). The ancilla initial squeezing is
5 dB (a),  9 dB (b), and 14 dB (c), which corresponds to $1/s=1.78$, 2.82, and 5.01, correspondingly.}
 \label{fig_Cats_bad}
 \end{center}
 \end{figure*}

\begin{figure*}[h]
 \begin{center}
 \includegraphics[width=0.9\textwidth]{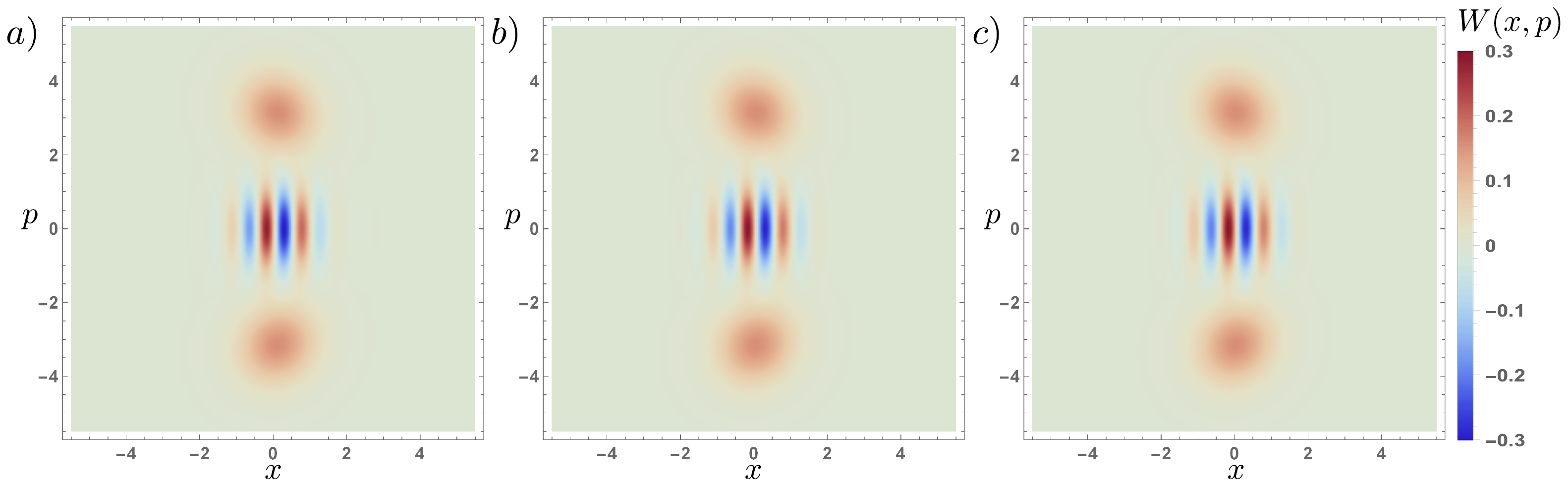}
 \caption{Same as in Fig. \ref{fig_Cats_bad}  for the cubic deformation coefficient $\gamma=0.5$ and the ancilla
momentum measurement outcome $y_m=15$ (high fidelity configuration).}
 \label{fig_Cats_good}
 \end{center}
 \end{figure*}

Fig. \ref{fig_Infidelity} shows that starting from some threshold squeezing, which is achievable in
experiment, a further increase in squeezing does not lead to a significant increase in fidelity. 
The fidelity difference from unity
in this limit is due to local deformations of the input state copies on the phase plane,
as seen from the Figs. \ref{fig_Cats_bad} and \ref{fig_Cats_good}. 
This deformation is the smaller, the larger the parameter $\gamma$ of cubic nonlinearity.

 
 \begin{figure}[h]
 \begin{center}
 \includegraphics[width=0.45\textwidth]{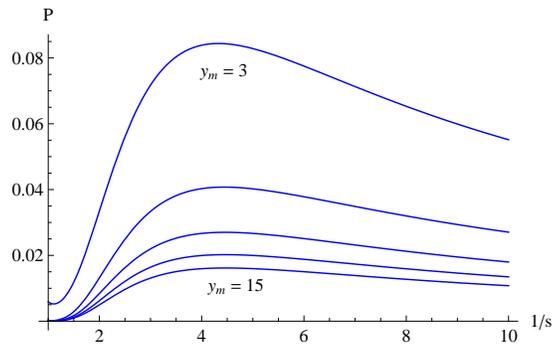}
 \caption{Probability density $P(y_m)$ of the measurement outcome $y_m=3,\,6,\,9,\,12,\,15$ respectively
 in dependence on the  initial ancilla state squeezing, where $1/s \geq 1$ is the stretching factor of the ancilla coordinate
 quadrature.  For any measurement outcome in these plots, the cubic deformation coefficient is assumed to 
 be $\gamma = y_m/30$, which provides  the spacing $2p^{(+)} = 3.16$  between the centers of copies 
 along the momentum axis. }
 \label{fig_Probability}
 \end{center}
 \end{figure}

\begin{figure}[h]
 \begin{center}
 \includegraphics[width=0.35\textwidth]{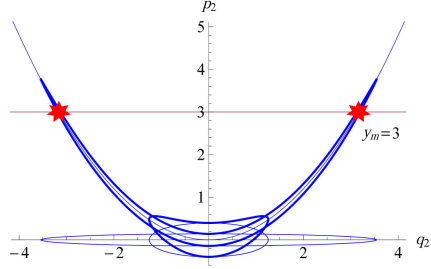}
 \caption{A visual representation of the ancillary oscillator support region on the phase plane before and after
 the cubic deformation with $\gamma=0.1$ for the initial squeezing of 5 and 14 dB respectively. 
Two distinct areas on the phase plane of the ancilla observables compatible
with the same momentum measurement outcome $y_m=3$ are  marked by the asterisks.
Due to entanglement, both values of the ancilla coordinate are transferred to the target oscillator
observables, thus generating the two-component cat state.}
 \label{fig_Ellipses}
 \end{center}
 \end{figure}

However, as seen from Fig. \ref{fig_Probability}, the degree of initial squeezing significantly affects the 
probability to obtain the expected measurement result $y_m$. This is due to the fact that the probability density 
per unit momentum interval of the ancillary oscillator, initially prepared in cubic state, essentially depends on 
squeezing, as illustrated in Fig. \ref{fig_Ellipses}.

For a qualitative interpretation of this feature of the gate under consideration,  we
introduce in Fig. \ref{fig_Ellipses} the support regions of cubic phase state for 5 and 14 dB squeezing, built
in the semiclassical approximation, i. e. when the evolution of points on the phase plane caused by
the cubic Hamiltonian $ -\gamma q_2^3$, is described via a semiclassical mapping of the form
$x_2 \to x_2, \quad y_2 \to y_2 + 3\gamma q_2^2$.

At low squeezing, the resource quasi-cubic state has a low probability density near $p_2 = y_m$,
as shown in Fig. \ref{fig_Ellipses} for 5 dB squeezing. For an arbitrarily large squeezing, the support region
of the resource state is stretched infinitely into an arbitrarily ``thin" parabola,
and the probability of the expected momentum measurement result vanishes.

\begin{figure*}[h]
 \begin{center}
 \includegraphics[width=0.7\textwidth]{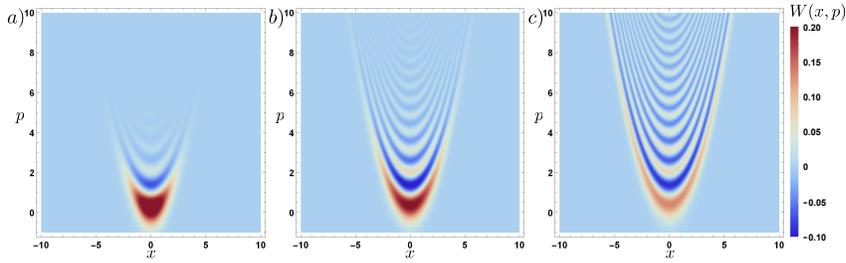}
 \caption{Wigner function of the non-ideal cubic phase state for the cubic deformation coefficient $\gamma=0.1$, 
 generated from the initial ancilla  state with squeezing of 5 dB (a), 9 dB (b)  and 14 dB (c).}
 \label{fig_Cubic_Wigner}
 \end{center}
 \end{figure*}

As is well known, a correct quantum description of the statistical distribution, which corresponds to
a given quantum state,  is represented by its Wigner function. In Fig. \ref{fig_Cubic_Wigner}
we show the Wigner function of the cubic phase state built for the same values of the
parameters of cubic interaction and squeezing, as in the figures \ref{fig_Cats_bad},
\ref{fig_Cats_good}, and \ref{fig_Ellipses}.

One can note a good agreement between the results that follow from the exact form (\ref{out}) of the 
gate output state, and the qualitative picture introduced above.
This means that despite a more complex behavior of the cubic phase state Wigner function in comparison  
with its semiclassical interpretation, some features present in the behavior of the Wigner function
(e.~g., oscillations) may average out and not manifest themselves in the output state.

Hence, in turn, we can conclude that our approach \cite{Sokolov20,Masalaeva22} to qualitative analysis
of the gates that use non-Gaussian resource states and measurement can be useful for the search 
and evaluation of other logical devices.


 \section{Efficiency of the gate versus monotones}

Recently, the problem of universal measures of non-Gaussianity of the states that can serve as a 
resource for universal quantum computing has been widely discussed \cite{Walshaers21}. One could expect that
a non-Gaussian resource state, for which a well-chosen measure is larger, would be able
to  take more efficiently the result of  quantum evolution out of the class of Gaussian schemes
when used optimally.

Among the measures discussed so far in this context are the non-Gaussianity and Wigner logarithmic negativity.
For finite squeezing, the non-Gaussianity and Wigner logarithmic negativity of the cubic phase state,
which our gate uses as a non-Gaussian resource,
are \cite{Albarelli18} monotonous functions of the parameter $\gamma /s^3$ (in the notation of our work,
$1/s \geq 1$ is the stretching factor of the coordinate quadrature). This parameter increases with the 
degree of squeezing. 

Our results demonstrate that for large enough parameter $\gamma$  of the cubic non-linearity, our gate can conditionally 
produce the output state which is very close to the cat-like superposition  (\ref{cat_Glauber_new}) 
of the Glauber states almost independently on the degree of squeezing 
(after some threshold value, see Fig. \ref{fig_Infidelity}). 
If the measured ancilla momentum $y_m$ fits to the range specified before the measurement,
the ``amount of non-Gaussianity''   in the output cat-like state  becomes, for large enough squeezing, almost 
independent on the squeezing present in the parameter  $\gamma /s^3$, no matter which measure is used.

However, from the point of view of a real use of the measurement-assisted gates, the concept of efficiency
should include not only the quality of the prepared states, but also the probability of obtaining the necessary
measurement result. In our scheme, the ``amount of non-Gaussianity'' present in the output state,
weighted by its probability shown in Fig. \ref{fig_Probability}, vanishes in the limit of perfect squeezing.
 
That is, the use of cubic phase state with the larger value of the parameter $\gamma/s^3$ 
and, hence, with the larger non-Gaussianity and Wigner logarithmic negativity, may render the gate less efficient
in the sense mentioned above. 
 
One can conclude from this example that the non-Gaussianity measures based on global properties 
of the resource states may be generally not suitable for the schemes based on the entanglement and projective 
measurements. Physically speaking, the efficiency of such schemes may depend more on the resource state behavior
in the region of phase space, where the resource state overlaps with the state 
detected by the measuring device, than on its global properties.


 \section*{Conclusion}

We have investigated quantum statistical properties of two-component Schr{\"o}dinger cats 
created at the output of the conditional non-Gaussian logical gate that we have proposed previously.
The gate is based on quantum entanglement of the target oscillator with the ancillary one,
and the subsequent projective measurement of ancilla.
A key feature of the regime where cat-like states arise is that the measurement provides 
multivalued information about the target system physical variables.

As a non-Gaussian resource, we considered here cubic phase state generated from the initial squeezed state 
of ancilla with an arbitrary finite initial squeezing.

We showed that in the relevant range of parameters, an increase in the degree of initial squeezing
above some threshold range achievable in experiment has negligible effect on the fidelity
between the gate output state and the Schr{\"o}dinger cat state in the form of  superposition of two 
coherent Glauber states.
On the other hand, we demonstrated that the degree of initial squeezing essentially affects the
probability of the projective measurement outcome that leads to the desired output
state. This probability vanishes for both too small and too large degree of squeezing,
and reaches its maximum value at a certain degree of squeezing, which depends on the degree
of cubic non-linearity used in the preparation of the resource state and on the 
output state parameters. We provided a simple physical interpretation of these results using 
both the exact and semiclassical representation of the resource state properties and of the 
measurement procedure on the phase plane.

Based on our results, we also draw attention to the fact that such measures of non-Gaussianity of the
resource state, as Wigner logarithmic negativiy and non-Gaussianity, may be not applicable 
to assess the overall efficiency of non-Gaussian gates which use quantum entanglement and
subsequent projective measurement.

 


 \bibliography{}
 \bibliographystyle{plain}
 
 \end{document}